\pdfoutput=1
\documentclass[letterpaper]{article}

\usepackage[utf8]{inputenc}
\usepackage{aaai19}
\usepackage{times}
\usepackage{helvet}
\usepackage{courier}
\usepackage{url}
\usepackage{graphicx}
\usepackage{mathtools}

\begin{document}
\title{Forecasting Intra-Hour Imbalances \\in Electric Power Systems}
\author{
  Tárik S.~Salem,${}^{1}$ Karan Kathuria,${}^{2}$ Heri Ramampiaro,${}^{1}$ Helge Langseth${}^{1}$\\
  ${}^{1}$Norwegian University of Science and Technology (NTNU), Department of Computer Science, Trondheim, Norway\\
  ${}^{2}$Optimeering AS, Oslo, Norway\\
  tarik.salem@ntnu.no, karan.kathuria@optimeering.com, \{heri, helgel\}@ntnu.no
}
\maketitle
\begin{abstract}
Keeping the electricity production in balance with the actual demand is becoming a difficult and
expensive task in spite of an~involvement of experienced human operators. This is due to the
increasing complexity of the electric power grid system with the intermittent renewable production
as one of the contributors. A beforehand information about an occurring imbalance can help the
transmission system operator to adjust the production plans, and thus ensure a high security of
supply by reducing the use of costly balancing reserves, and consequently reduce undesirable
fluctuations of the 50 Hz power system frequency. In this paper, we introduce the relatively new
problem of an~intra-hour imbalance forecasting for the transmission system operator (TSO). We focus
on the use case of the Norwegian TSO, Statnett. We present a complementary imbalance forecasting
tool that is able to support the TSO in determining the trend of future imbalances, and show the
potential to proactively alleviate imbalances with a~higher accuracy compared to the contemporary
solution.
\end{abstract}

\section{Introduction}

The electrical power system is a highly complex system. One of the main reasons is the need for
instantaneous balancing of supply and demand in any synchronous power system. The Nordic power
system (consisting of the countries Norway, Sweden, Finland and Denmark) is no different.
Additionally, it is undergoing a~period of significant changes driven by the developments in the
infrastructure and consumer behaviour~\cite{StatnettEtAl-2016}. The volumes of intermittent
renewable capacity has increased rapidly over the past decade, driven by a mix of environmental
regulations and subsidies, and cost reduction. Furthermore, consumers of power are increasingly
generating their own power. The increased volatility in production and demand resulting from these
developments represent a significant challenge to the operation of power systems.

Also following the liberalisation of the Nordic power market in the 1990s~\cite{Glachant-2014}, many
of the tasks related to coordinating the supply and demand have been given to the market through the
creation of the Nord Pool\footnote{\url{https://www.nordpoolgroup.com/}}, a Nordic market for
electric power exchange. In this market place, Nordic market actors exchange bids and offers and
agree on the future production and delivery of physical power~\cite{Wangensteen-2012}. Nevertheless,
because of structural limitations in the market (mainly the time resolution being limited to hours)
and the uncertainty associated with both generation and demand, there will always be a need for a
transmission system operator (TSO) to handle instantaneous imbalances (i.e. differences between
demand and supply) within the operating hour. As imbalances ultimately result in frequency
deviations in the synchronous power system, managing these imbalances is a key in ensuring a stable
and secure power supply.

\begin{figure}[t]
  \centering
  \includegraphics[width=0.47\textwidth]{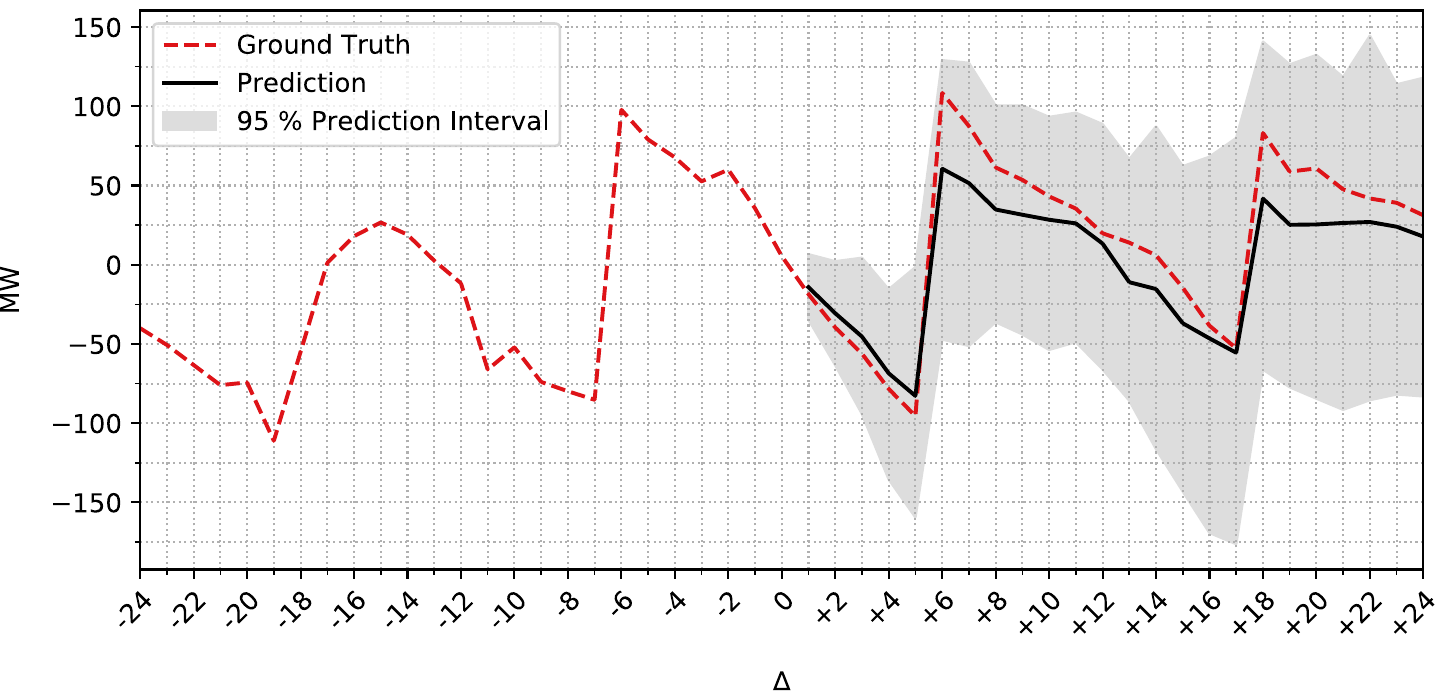}
  \caption{A two hours ahead imbalance prediction with prediction intervals.}
  \label{fig:prediction}
\end{figure}

The TSOs work tightly together (e.g. Statnett with their Nordic counterparts such as Svenska
Kraftnät, Energinet and Fingrid) to mitigate imbalances. Their toolbox for managing imbalances can
be dived into the following categories:

\begin{itemize}
    \item{
      \textit{Preventive measures.} These are measurements that are taken several hours
      before the operating hour, but after the market has determined the supply and demand. The
      TSO can then approximate the upcoming imbalances based on the supply and demand scheduled on
      an hourly resolution, and their intra-hour load forecast. The results of this approximation
      are presented in a so called Planning Table. Typically this results in smoothing out big
      hourly shifts in the scheduled production since the demand normally does not follow strict
      hourly patterns~\cite{NordicBalancingPhilosophy-2016}.
    }
    \item{
      \textit{Operational measures.} There are three types of balancing reserves that can be
      activated within the operation hour usually in the given order:
      \begin{itemize}
          \item{
          Frequency Containment Reserves (FCR) which is a built-in flexibility in the power system
          resulting from requirements that the TSO has set for all generators providing power to the
          system.
          }
          \item{
          Automatic Frequency Restoration Reserves (aFRR) which are automatically activated by the
          TSO and has a response time of maximum 2 minutes.
          }
          \item{
          Manual Frequency Restoration Reserves (mFRR) which can be ordered by the TSO manually by
          communication with large power producers and which is set to have a response time of
          maximum 15 minutes.
          }
      \end{itemize}
    }
\end{itemize}

In brief, the tools that are available for Statnett and other Nordic TSOs vary in terms of
activation time, response time, magnitude of availability and cost. As a balancing strategy,
Statnett uses mFRR both retrospectively (to free FCR and aFRR volumes that have been activated due
to an occurred imbalance) and proactively (to mitigate upcoming imbalances). The objective is to
minimize the use of expensive and limited FCR and aFRR, and thereby ensuring a high security of
supply~\cite{Haberg-2016}. In order to do so, Statnett relies on a human experience and the
aforementioned Planning Table, which provides forecasts for the upcoming imbalances in the next two
hours.

In essence, Planning Table is a prediction tool that takes the scheduled production from the market,
and subtracts a 5-minute granularity demand (load) forecasts to calculate the upcoming imbalance.
Unfortunately, the quality of the forecast provided by Planning Table is limited. Arguably, this is
because it relies greatly on the assumptions that the market is in balance on an hourly granularity,
that the generators produce the power without deviations from the plan, and that the demand
prediction is accurate. This lack of an effective tool for forecasting upcoming market imbalances
makes it hard to proactively manage imbalances with manual reserves, which in turn makes more use of
the expensive and limited aFRR and FCR. As a result, it becomes difficult for the TSO reaching their
KPI (Key Performance Indicator) in terms of frequency quality (i.e. minutes of frequency deviations
in the past years) and experience increasing costs of balancing \cite{StatnettEtAl-2016}. This,
coupled with the fact that published literature on forecasting system imbalance volumes is very
limited, indicates great room for improvement~\cite{Klaebu-2013}, and is the main motivation behind
this research.

In order to mitigate imbalances proactively and effectively, an improved or complementary
forecasting tool would be of a great benefit to TSOs like Statnett. Such a tool would additionally
implement some of the expert knowledge, and thus lessen the burden of the human experts, improve the
decisions and the robustness of the power system, together with reduction of the costs for
balancing.

In this work, we present a complementary forecasting solution giving the TSO information about the
most likely trend of imbalances within the next two hours. Additionally, it is beneficial to have an
information about the reliability of predictions especially in use cases where human experts are
involved in the process of decision making. For this reason, the aforementioned imbalance
predictions are accompanied with prediction intervals. By applying quantile forest regression on
feature engineered data, our solution outperforms a proxy of Statnett's contemporary forecasting
solution, i.e. a more accurate representation of the Planning Table.

\section{Problem Definition}

We define the imbalances as a time series with 5-minute granularity, where each value represents an
average imbalance of the past five minutes. The objective is to predict the imbalances two hours
ahead. Formally, in a market area $\alpha$ and at time $t$, our objective is to predict 24 future
imbalances $I_{\alpha, t+\Delta}$, where $\Delta \in \langle1, 24\rangle$, alternatively annotated
as $\boldsymbol{I_{\alpha, t}} = (I_{\alpha, t+1}, ..., I_{\alpha, t+24})$. Imbalance predictions
are denoted with $\hat{ }$, i.e. $\hat{I}_{\alpha, t+\Delta}$ or $\boldsymbol{\hat{I}_{\alpha, t}}$.
The goal is to minimize the difference between the predicted imbalance $\hat{I}_{\alpha, t+\Delta}$
and actual imbalance $I_{\alpha, t+\Delta}$. Additionally, each imbalance prediction
$\hat{I}_{\alpha, t+\Delta}$ shall be accompanied with prediction interval $PI_{\alpha, t+\Delta}$
which is an estimate of an interval where an observed imbalance, i.e. $I_{\alpha, t+\Delta}$, will
lie with a certain probability. In this work, $\alpha$ is one of the five market
areas (labeled as $NO1$, $NO2$, $NO3$, $NO4$ and $NO5$) into which Norway is currently divided.

A key aspect is the definition of what precisely a single imbalance value represents. Since all
imbalance managing measures share the purpose of reducing the imbalance, the imbalance itself varies
with respect to which point in time it is measured. As the target model is intended to be an
operational decision support tool for activating operational measurements, the imbalance is defined
as the imbalance measured after preventive measures is taken, but before any operational measures
have affected the imbalance. The imbalance is measured in megawatts (MW) and is commonly referred to
as Area Control Error Open Loop (ACE OL). This definition ensures that the balancing measures (taken
based on the predictions) do not affect the imbalances. Worth underlining is also that imbalance
(ACE OL) can be computed only retrospectively at the end of the period of interest, as it
essentially is determined by subtracting the actual operational measures from the actual observed
imbalance (so called Area Control Error, or ACE). Additionally, the preventive measures are
subtracted from ACE OL as they are planned ahead and known prior the prediction time
\cite{NordicBalancingPhilosophy-2016}.

\begin{figure}[tp]
  \centering
  \includegraphics[width=0.47\textwidth]{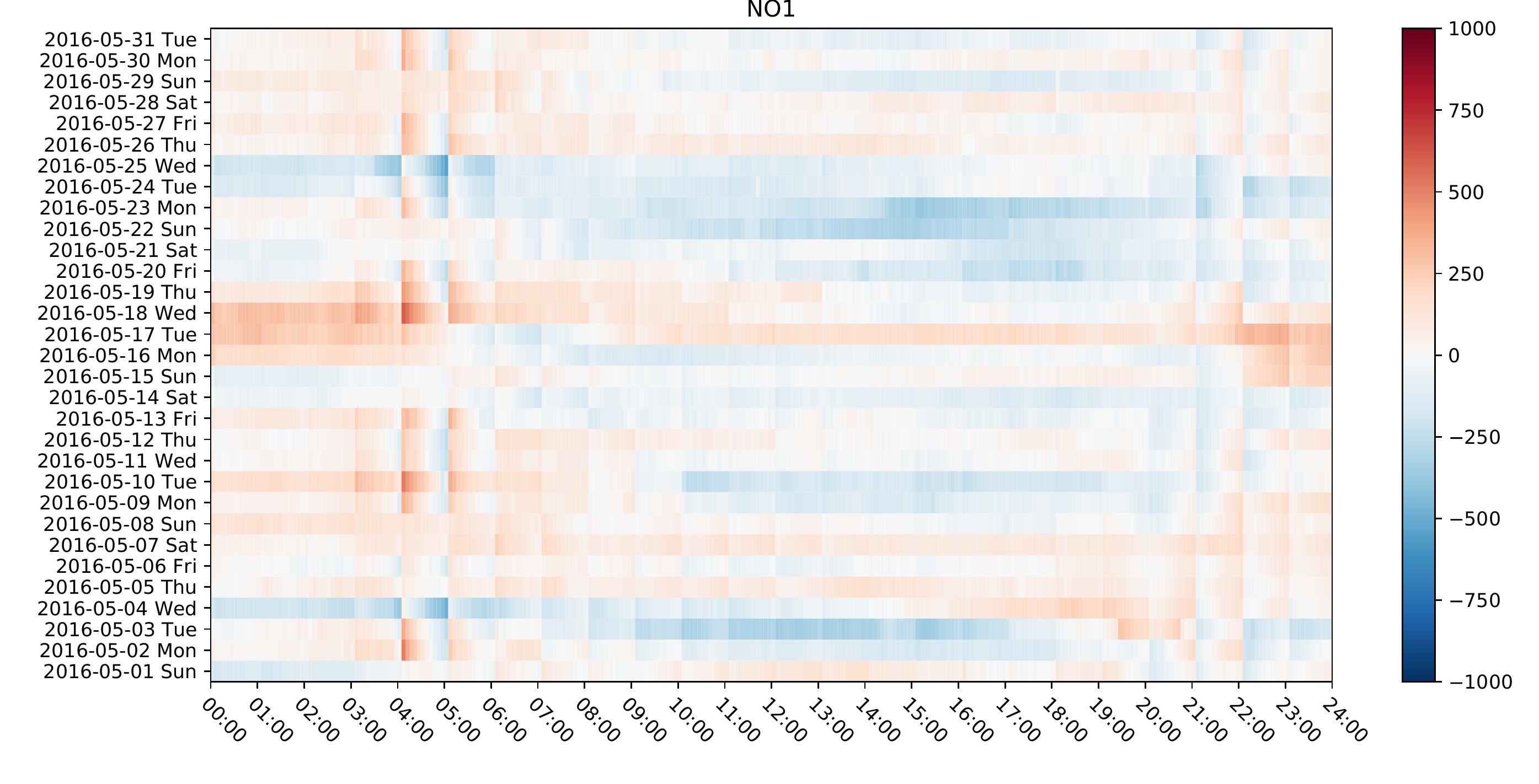}
  \includegraphics[width=0.47\textwidth]{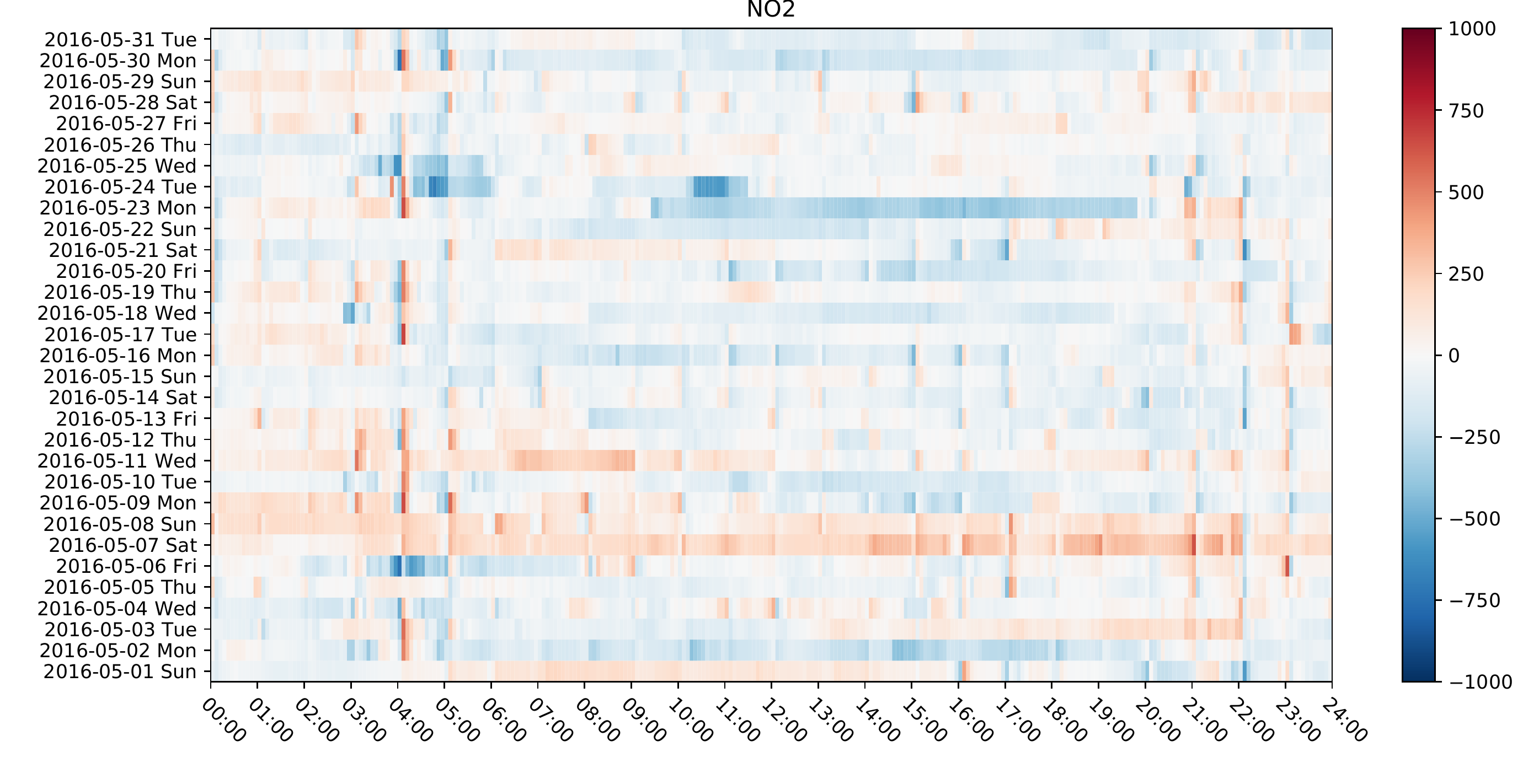}
  \includegraphics[width=0.47\textwidth]{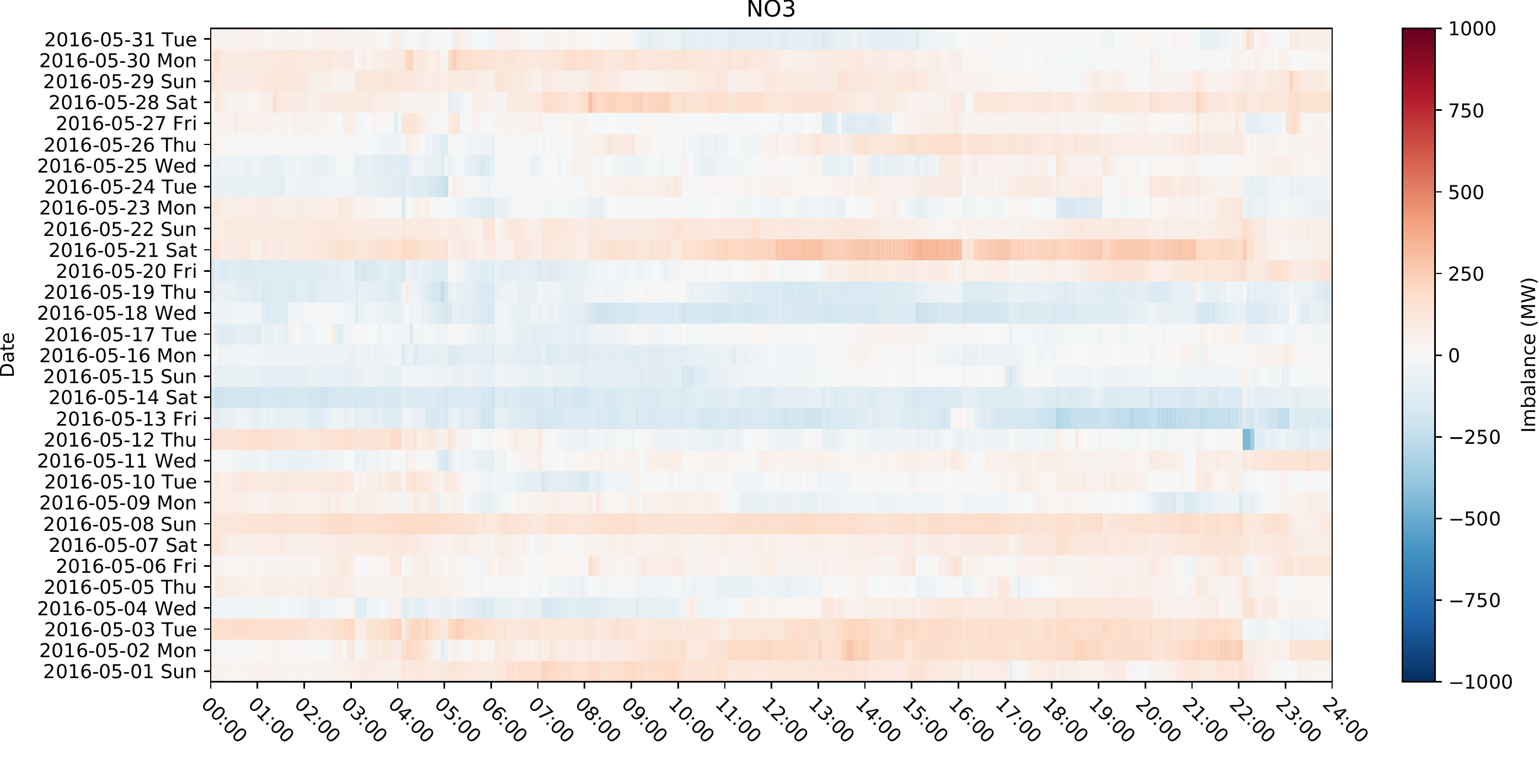}
  \includegraphics[width=0.47\textwidth]{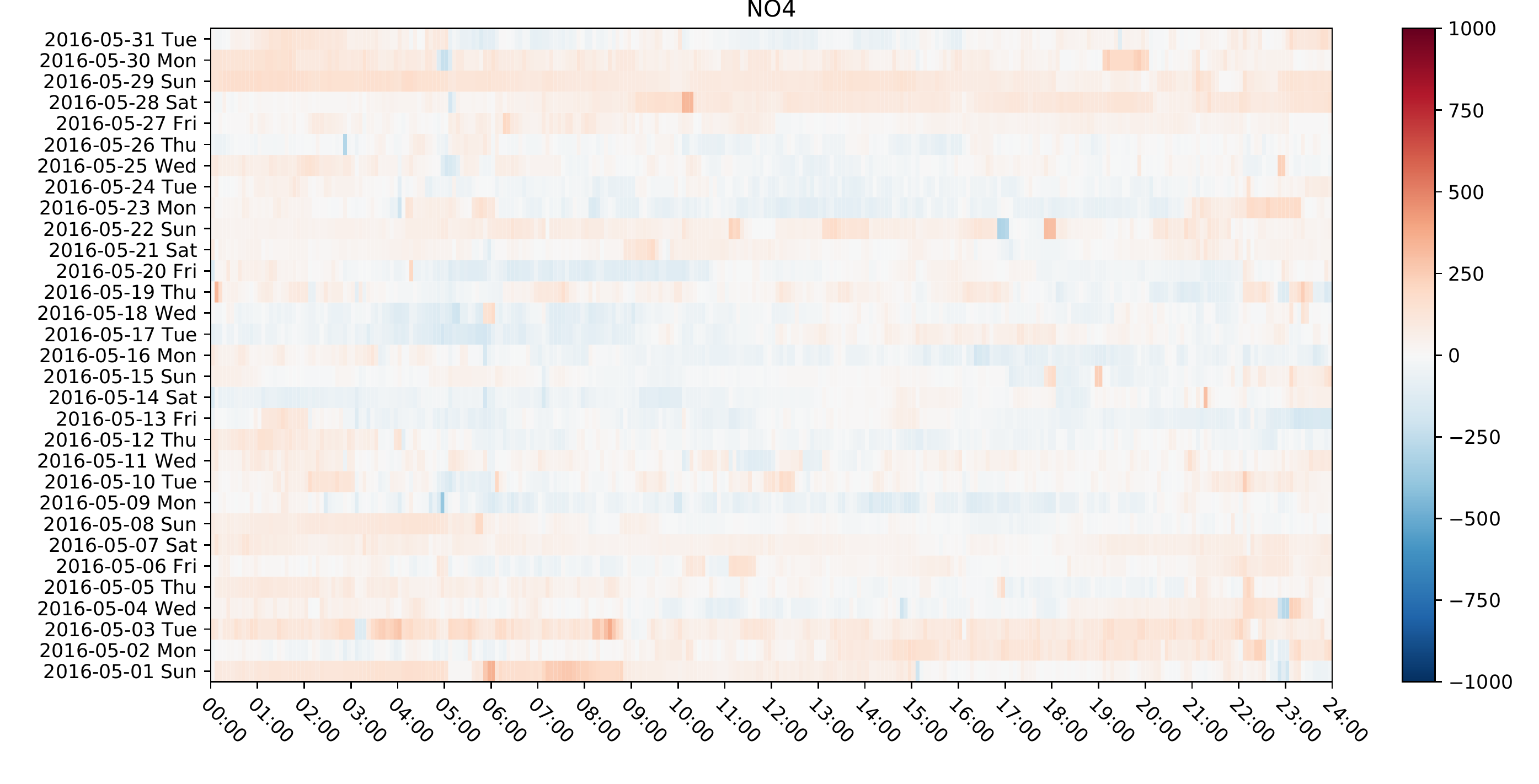}
  \includegraphics[width=0.47\textwidth]{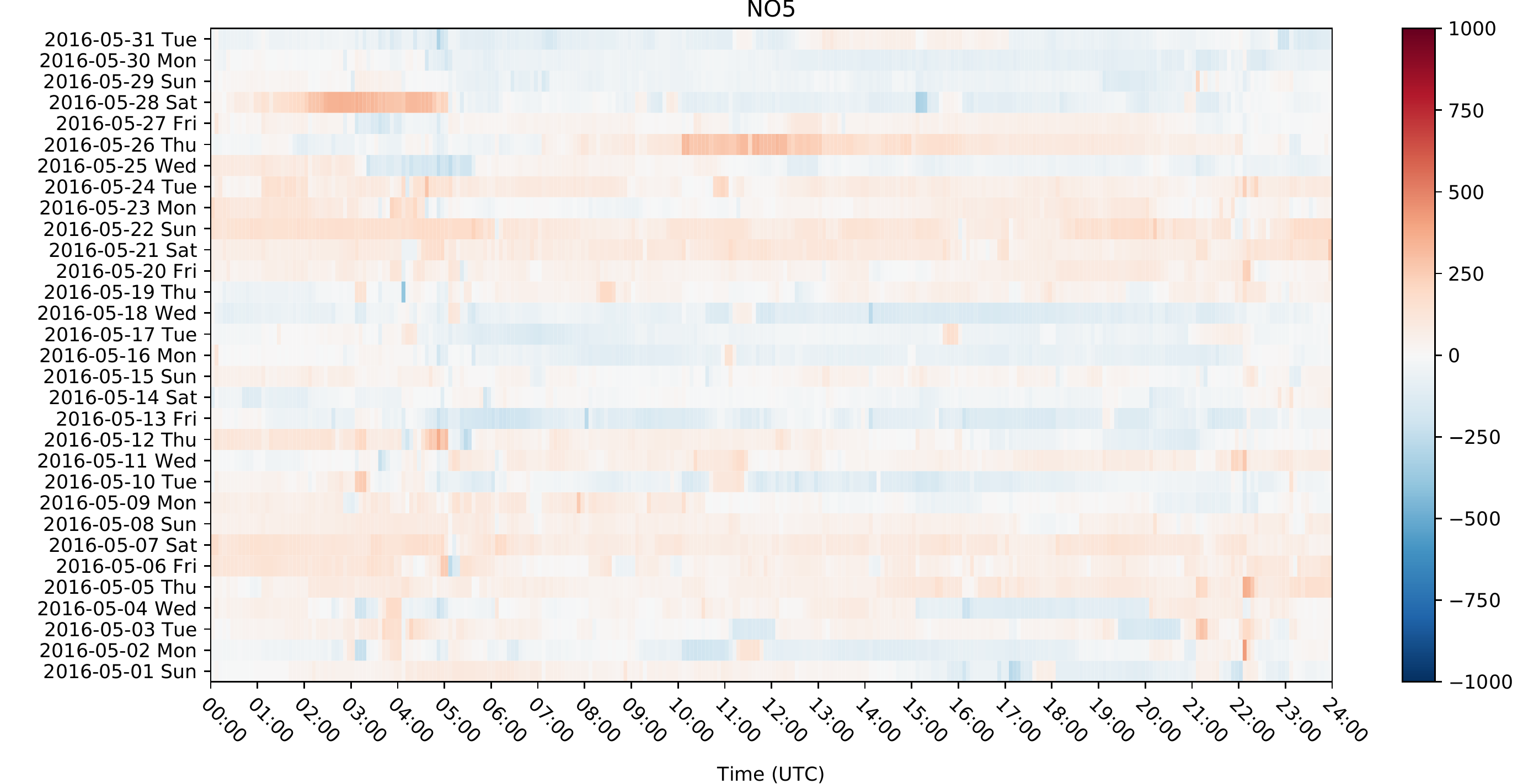}
  \caption{Imbalances in the Norwegian market areas in May 2016.}
  \label{fig:imbalances-detailed}
\end{figure}

Figure \ref{fig:imbalances-detailed} visualizes in a great detail the heterogeneous character of the
pre-processed imbalances in the market areas within the same timespan (i.e. May 2016).
Apparent is a strong correlation over time and differences between the areas. Notice the
sudden changes in imbalance at the beginning of hours (e.g. in area NO1 at 3, 4 and 5 o'clock in the
morning). This is due to the aforementioned hourly resolution of the planned production contrary to
the abrupt intra-hour changes of the demand, even though possibly matching the hourly average of the
production.

\section{Related Work}

To the best of our knowledge, only a handful of publications
\cite{garcia:2004,kratochvil:thesis:2016,contreras:thesis:2016} are dealing with the power system
imbalance forecasting, which gives a hint about the novelty of the topic. In comparison, a large
number of publications deal, e.g., with the short-term load forecasting, prediction of market
prices, and analysis of the impacts of increasing penetration of intermittent renewable sources. The
following provides a brief overview of the three publications dealing with power system imbalance
predictions.

The first publication \cite{garcia:2004} discusses and shows the limitations and insufficiency of
richly applied but basic forecasting techniques such as ARIMA and exponential smoothing, because
of the non-periodic, non-stationary and noisy character of imbalance time-series. Instead, the
contemporary state-of-the-art artificial neural networks (ANN) are applied to uncover the
non-linearity and irregularity of the data, and predict the daily imbalance medians. Presented are
improvements compared to methods based on a linear regression, and the fact that none of the neural
networks provided optimal predictions for all market conditions is discussed. Two use cases were
evaluated: prediction of daily medians with three months training and one month testing window, and
prediction of six values for each day with four week training and one week testing window. The
following predictor variables were employed: demand forecast, demand forecast error, accepted bid
volumes, accepted offer volumes, forward trades, gate closure imbalance volume, accepted offers and
bids, imbalance prices and day of the week. However, it is not clear whether historical imbalances
were utilized as one of the input features of the models.

The second publication \cite{kratochvil:thesis:2016} investigates the most important predictors
contributing to accurate forecasts. The impact of multiple exogenous variables was analyzed by
applying autocorrelation analysis. The exogenous variables were grouped into three classes: demand
variables, supply variables, and market participants' behaviour. One of the outcomes of the analysis
is that variables of maximum two hours delay provide sufficient predictive power. However, we should
be aware of the fact that autocorrelation is the linear dependence of a variable. The proposed model
does not aim to predict imbalances accurately, instead, its objective is to map the future imbalance
into one of five proposed intervals. The input variables have undergone similar discretization. The
discretized dataset is used in a customized linear-regression-based model for forecasting intra-hour
(with one minute granularity) imbalances. The model takes into account the exogenous variables and
is being compared to the ARIMA based benchmark. Unfortunately, the current state-of-the-art
modelling approaches do not seem to be exploited in this work.

The third publication \cite{contreras:thesis:2016} is dedicated to hourly imbalance predictions
using random forest regression. Similarly to previous works, a broad range of influencing variables
are taken into account. The listed variables may be grouped into demand forecast, wind forecast,
temperature, bid volume in day-ahead market, day-ahead market price, energy volume auctioned in
phase 2 of technical constraints and its price, reserves. The final model is trained with only
eight-day dataset, and evaluated on the consequent ninth day. Multiple variants of feature sets were
compared of their predictive performance. Model incorporating all features performed best, however,
with a small margin. The work also presents a variable analysis predominantly exploring linear
dependencies. However, it is not clear how the 24 target imbalances were predicted using random
forests. It is worth noting that historical imbalances were not included in the feature set.

Incorporating the exogenous variables is an important research direction, and all of the above
mentioned works have this in common. Their incorporation of influencing variables is related to
several factors, including demand forecasts, market prices, planned production, weather observations
and forecasts.

Of the aforementioned works, only the work by~\citeauthor{kratochvil:thesis:2016}
(\citeyear{kratochvil:thesis:2016}) copes with the problem of our interest, i.e. short-term power
imbalance forecasting. Still, the aims of our solution are slightly different. The proposed solution
should not replace the contemporary tool, Planning Table, but rather act as a complementing tool
enhancing the judgment of the human experts. This is also due to existing constraints regarding the
available data sources (as described in the Dataset section). Another important distinction to the
mentioned works is that our solution provides prediction intervals, in addition to point predictions
of imbalances.

\section{Forecasting Intra-Hour Imbalances}

This section describes the selected modelling approach, i.e. quantile regression
forests~\cite{Meinshausen-2006}, and the experimental setting producing the presented results.

The quantile regression forests (QRF) is a generalization of the random forests~\cite{Breiman-2001},
a~sound and widely applied ensemble learning method. Random forests (RF) provide point predictions
in the form of conditional mean $E(Y|X = x)$, where $Y$ is the target variable and $X$ the predictor
variable. QRF extends RF with the ability to generate prediction intervals as an intrinsic
part of the model, thus allowing to assess the reliability of the predictions. Given $X = x$, the
prediction interval $\mathcal{Q}_{p}(x)$ expresses that a new observation Y will lie with a certain
probability ($p$) within an interval given by the lower quantile $Q_{l}$ and upper quantile $Q_{u}$.
For example, a 95~\% prediction interval is given by
\begin{equation*}
\mathcal{Q}_{.95}(x) = [Q_{.025}(x), Q_{.975}(x)].
\end{equation*}
\noindent
The $q$-quantile is then defined as
\begin{align*}
Q_{q}(x) = \inf\{y : P(Y \leq y|X = x) \geq q\}.
\end{align*}
In RF (and QRF), the conditional mean $E(Y|X = x)$ is approximated as a weighted mean over the
observations of the target variable $Y$. Similarly in QRF, the conditional distribution function
$P(Y \leq y|X = x)$, defined as $E(1_{\{Y \leq y\}}|X = x)$, is approximated by weighted mean over
the observations of $1_{\{Y \leq y\}}$. Therefore, each leaf of each tree stores also the
observed target variables $Y$ besides their average.

The prediction of the 24 future imbalances is treated as an univariate prediction problem. This
means that for each of the 24 predicted imbalance values, an independent model is trained. The
imbalances are predicted for each price area separately. This results in 120 models in the given
five market prices areas.

Through an empirical testing of training set sizes ranging from 3 to 18 months, the
training sets of size at least 12 months manifested considerable performance gains. Greater
training set sizes did not show evidence of diminishing performance. Therefore, the training set
used in this paper consists of one year of data. The 12 months training set size was chosen also
with regards of the test set size, allowing us to analyze the performance throughout the year. Also
through empirical testing we have selected 1 month as a retraining interval, i.e. once per month a
new model is trained with the preceding 12 months as training set. Note, however, that it is beyond
the scope of this paper to present the empirical results for selecting the sizes of training set and
test set.

Through an exhaustive hyper-parameter search with the focus on mean squared error (MSE), we were
able to select hyper-parameters for each of the predicted imbalance values ($I_{\alpha, t+\Delta}$).
However, as the gains were not significant, we selected one hyper-parameter set applied across all
experiments. Here, the key hyper-parameters are 100 estimators (with minimum leaf size set to ten
observations), each trained on bootstrapped data retaining a subset of 70~\% randomly selected
features. In our experiments, MSE was used as the splitting criterion.

Easy maintenance, i.e. retraining and adjustments in features, plays a key role when
selecting the model type. RF type of models are known for their non-sensitiveness to overfitting and
noise, and our empirical testing has also shown a small sensitivity to changes in HPs, i.e. small
changes in the HPs do not lead to considerably different results. It is worth noting that QRF had a
decisive influence on the model selection.

\section{Dataset}
\label{sec:dataset}

This section describes the available data (defining the scope and limitations of this work), and the
dataset used for training and testing of the proposed models. Consequently, we clarify the
definition of an imbalance, a key aspect of the problem definition.

We base this work on a dataset provided by Statnett, which contains two years of imbalance data from
2015 to 2016. The dataset individually covers the five Norwegian market areas (pricing areas),
recalling that the focus is on predicting imbalances in each market area separately. We will refer
to the market areas as NO1, NO2, NO3, NO4 and NO5. The imbalance data is a time series made up by
continuous timestamped values with a 5-minute granularity, where each value represent the average
imbalance across the past 5 minutes ending at the timestamp. For each market area, the dataset
comprises roughly 210K data points, in total.

The imbalance can be decomposed into three main contributing factors: error of the consumption
forecast, deviations from the production plan, and deviations from the HVDC plan (i.e. planned
import and export of the power). Here we ignore unforeseen events, such as failures in the
electrical power grid, as another contributing factor. Each of the factors is further under
influence of exogenous variables such as weather and human activity. Unfortunately, we were not able
to decompose and utilize the imbalance contributors, due to the absence of the data. For this
reason, we limit the scope of this work to following types of variables: historical imbalances,
temporal features (including solar elevation), market prices, planned HVDC flows, production plans,
and forecasts and observations of the temperature, wind and river flows.

The final dataset is a result of an exhaustive manual feature engineering (partly involving human
experts from Statnett and Svenska Kraftnät) and feature selection process. We have
evaluated the top-N features retrieved by the feature importance ranking of the RF. Additionally,
aware of the feature importance diffusion of correlated features, multiple manually created (using
previous results, the domain expertise and logical reasoning) feature subsets were evaluated.
However, due to space constraints of this paper, we only present the final and simplest version of
the utilized dataset. The final feature set, acting as predictor variables ($X$) in our
model, has exhibited greatest gains in terms of predictive performance. The key features were
dramatically reduced to the following: 24 past imbalances ($I_{\alpha, t - \delta}$, where $\delta
\in \langle1, 24\rangle$) represented as relative values to $I_{\alpha, t + 0}$, $I_{\alpha, t + 0}$
(as an absolute value), and temporal features (month, day of the week, hour, minute, sine and cosine
representation of the previously listed, holiday flag, and solar elevation). The prediction targets
are also imbalance values relative to $I_{\alpha, t + 0}$.

\begin{figure}[t]
  \centering
  \includegraphics[width=0.47\textwidth]{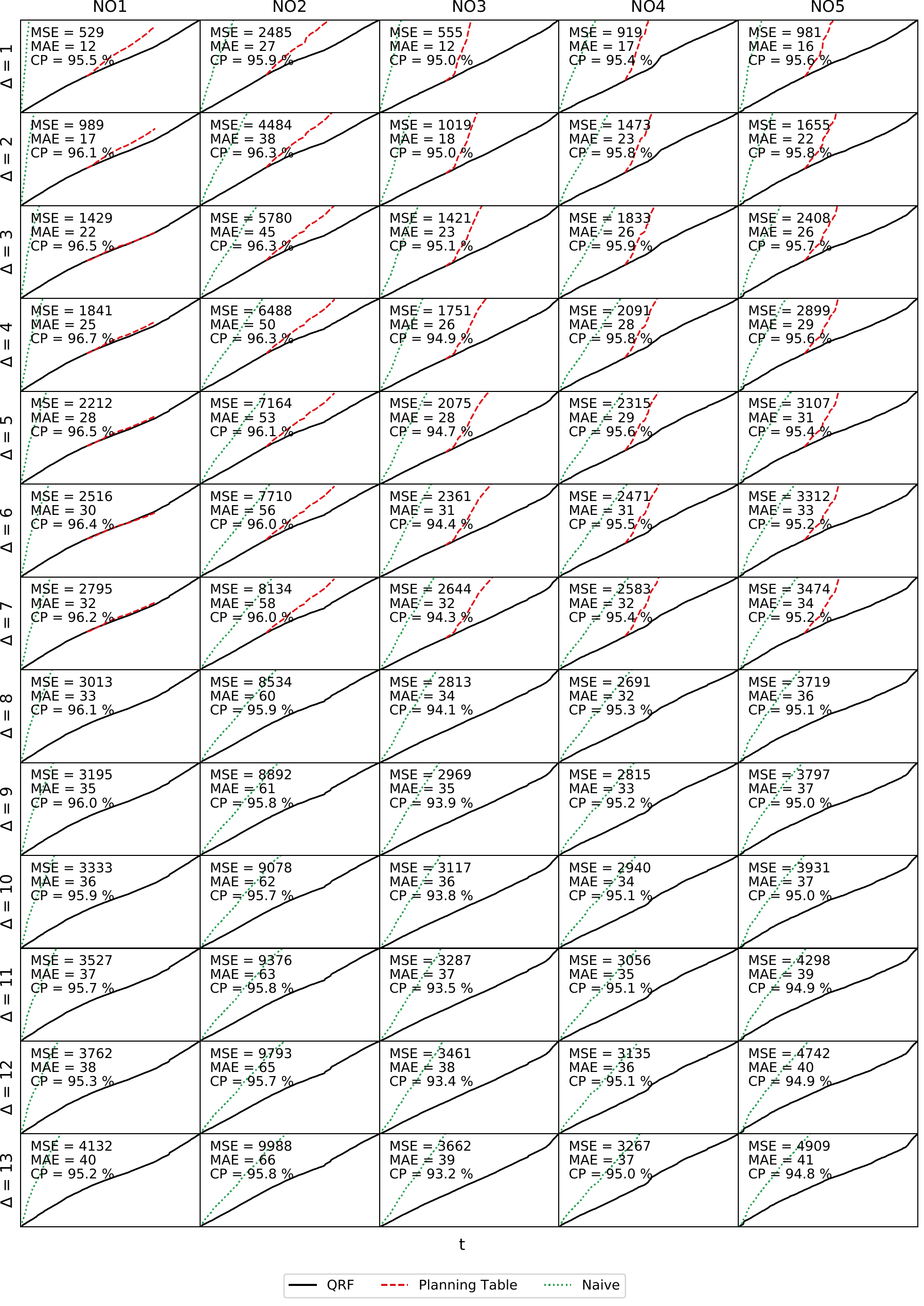}
  \caption{Plots of cumulative errors over time $t$ for each market area $\alpha$ and
  prediction target $t+\Delta$ in the year 2016. Due to space restrictions only results of
  $\Delta \in \langle1, 13\rangle$ are presented. The Planning Table benchmark is available only for
  a subset of data, i.e. for $\Delta \in \langle1, 7\rangle$ from March to September 2016.}
  \label{fig:results}
\end{figure}

\section{Experimental Results}

The predictions were evaluated on the period from January to December 2016. Each model was trained
on 12 calendar months, and evaluated on the subsequent calendar month. This results into a set of 12
prediction sets times 24 predicted variables (two hours of imbalances) times 5 market areas.

For point predictions, we calculate the mean squared error (MSE) and the mean absolute error (MAE)
to measure the accuracy. To evaluate the prediction intervals, we calculate the coverage probability
(CP), which is essentially the probability that the future (observed) imbalances will lie within the
constructed prediction intervals, and is given by\\
\begin{equation*}
CP = \frac{1}{N}\sum_{i=1}^{N} c_{i}
\quad\text{where}\quad
c_{i} =
\begin{cases}
  1 & \text{if}\ y_{i} \in \mathcal{Q}_{p}(x_{i}); \\
  0 & \text{otherwise},
\end{cases}
\end{equation*}
\noindent
where $N$ is the total number of predictions, $y_{i}$ is an observed value and $\mathcal{Q}(x_{i})$
is the prediction interval. The 95~\% prediction intervals were evaluated.

Two benchmarks are provided for the point predictions. The first benchmark, referenced as
\textit{Naive}, is a predictor assuming imbalances to be identical to previous week's. The second
benchmark, referenced as \textit{Planning Table}, is a more accurate proxy of the Planning Table.
This is due to missing consumption forecasts being replaced by actual consumption. Note that the
Planning Table benchmark is available only for $\Delta \in \langle1, 7\rangle$ from March to
September 2016.

Figure~\ref{fig:results} summarizes the results of the described experiments. Each cell plots
the cumulative error throughout the year 2016 for the QRF forecasts and benchmarks. Additionally,
each cell contains information about the MSE, MAE and CP of the QRF predictions. Each cell stands
for a specific prediction target given by market area $\alpha$ and time $t+\Delta$.

Above all, the predicted imbalances are of higher accuracy compared to both benchmarks. As expected,
the accuracy of the point predictions declines going further in the future (i.e. with increasing
$\Delta$). As desired, the coverage probability is at approximately 95 \% across all predictions.
Significant is the performance difference for the market areas NO2, NO3, NO4 and NO5, where the QRF
predictions are superior compared to the Planning Table.

We use the Scikit Garden\footnote{https://github.com/scikit-garden/scikit-garden}
implementation of the quantile regression forests.

\section{Prototype}

Although the quantitative offline evaluation showed considerable improvements in
forecasting imbalances, the actual prototype deployment (and usage by human operators) will
demonstrate the usefulness as an accompanying tool next to the Planning Table.

The lower MSE values can be translated to smaller frequency deviations, compared to the
Planning Table, if both solution would act as an autonomous system. However, neither of the two are
autonomous systems, as already mentioned they act as decision support tools. The human operator has
the final decision on the operational measures to take. Hence, after the deployment, we expect to
acquire a qualitative feedback from the human operators.

It is worth noting that the research and the deployment of the prototype would not be possible
without a tight cooperation with the TSO, i.e. Statnett. Normally, TSOs has conservative and careful approach in
introducing new tools in the control room (to the human operators). Nevertheless, the introduced
tool cannot (for reasons explained later) and does not aim to replace the Planning Table. Planning
Table still remains the main decision support tool, whereas our tool aims to provide more accurate
intra-hour imbalance forecasts utilizing historical data.

\section{Discussion}

The presented approach has its limitations. Contrary to the Planning Table, the model
does not incorporate features such as production plans, consumption forecasts and planned HVDC flows.
Thus, it cannot account for rare but foreseen imbalances. This is why the presented approach
cannot replace the Planning Table. For future work, it is crucial to take plans and forecasts into
account, and learn (reoccurring) patterns.

Additionally, we focus on the problem within the Nordic area. The imbalances 
may have different character in other electrical power systems, and thus our approach can exhibit
better or worse performance, and may require a different feature set.

The presented model does not utilize weather information because those features do not
contribute to substantial improvements. We hypothesise that the weather data may provide useful
information if the model could incorporate the consumption forecasts, or in electrical power systems
with a greater share of intermittent renewable production (e.g. wind and solar power).

\section{Conclusion}

In this paper, we proposed a method that applies quantile regression forests to predict two hours
ahead imbalances in the form of point predictions and more importantly as prediction intervals. In
contrast to known methods reported in the literature, the proposed method relies purely on
historical imbalances and features related to date and time, making our method novel for imbalance
forecast in general.

An unanticipated finding was that the imbalances exhibit relatively strong and to some extent
predictable patterns. Moreover, even though our model operates purely using historical imbalances
and temporal features, it was still able to outperform the contemporary solution of the TSO.

Overall, we see this work as an initial step towards a decision support tool, which could in a
longer horizon be a fully automated solution that is able to replace the insufficient Planning Table
altogether with the human expertise. For future work, we plan to develop a model that incorporates
some essential predictors such as consumption forecasts. More importantly, we are in the process of
building a prototype system that is able to deliver real-time imbalance forecasts. In addition to
being a proof of concept, this prototype system will be deployed in Statnett's control room and thus
provide a complementary tool for the human operators to assess the trend of the upcoming imbalances.
Furthermore, we expect to acquire a qualitative feedback from the human operators, which
could help us to further improve our method.

\section{Acknowledgements}

The project is funded by the Research Council of Norway, Statnett and Optimeering, and was carried
out by the Norwegian Open AI Lab at NTNU and Optimeering.

\bibliography{main-bibliography}
\bibliographystyle{aaai}

\end{document}